\documentclass[useAMS]{mn2e}
\usepackage{times}
\usepackage{graphicx}

\pubyear{2004}

\author[Bagla and Ray]
{J.S.Bagla and Suryadeep Ray \\
  Harish-Chandra Research Institute,  Chhatnag Road, Jhusi, Allahabad 211019,
  India. \\
  E-mail: (jasjeet, surya)@mri.ernet.in} 

\title[Size of the simulation box in cosmological N-Body simulations]
{Comments on the size of the simulation box in cosmological N-Body simulations}

\label{firstpage}

\def\LaTeX{L\kern-.36em\raise.3ex\hbox{a}\kern-.15em
    T\kern-.1667em\lower.7ex\hbox{E}\kern-.125emX}

\pagerange{\pageref{firstpage}--\pageref{lastpage}} \pubyear{2004}

\begin{document}

\maketitle


\begin{abstract}
N-Body simulations are a very important tool in the study of formation of
large scale structures.  
Much of the progress in understanding the physics of high redshift universe
and comparison with observations would not have been possible without N-Body 
simulations. 
Given the importance of this tool, it is essential to understand its
limitations as ignoring the limitations can easily lead to interesting but
unreliable results. 
In this paper we study the limitations arising out of the finite size of
simulation volume. 
This finite size implies that modes larger than the size of the simulation
volume are ignored and a truncated power spectrum is simulated. 
If the simulation volume is large enough then the mass in collapsed haloes
expected from the full power spectrum and from the truncated power spectrum
should match. 
We propose a quantitative measure based on this approach that allows us to
compute the minimum box size for an N-Body simulation. 
We find that the required box size for simulations of $\Lambda$CDM model at
high redshifts is much larger than is typically used. 
We can also use this approach to quantify the effect of perturbations at large
scales for power law models and we find that if we fix the scale of
non-linearity, the required box size becomes very large as the index becomes
small. 
The appropriate box size computed using this approach is also an appropriate
choice for the transition scale when tools like {\sl MAP}
\cite{1996ApJ...472...14T} that add the contribution of the missing power are
used.   
\end{abstract}




\section{Introduction}

Large scale structures like galaxies and clusters of galaxies are
believed to have formed by gravitational amplification of small
perturbations
\cite{1980lssu.book.....P,1999coph.book.....P,2002tagc.book.....P,2002PhR...367....1B}.   
Observations suggest that the initial density perturbations were present at
all scales that have been probed by observations.  
An essential part of the study of formation of galaxies and other large scale
structures is thus the evolution of density perturbations for such initial
conditions.  
The basic equations for this have been known for a long time
\cite{1974A&A....32..391P} and these equations are easy to solve when the
amplitude of perturbations is small.  
Once the amplitude of perturbations at relevant scales becomes large, i.e.,
$\delta \sim 1$, the perturbation becomes non-linear and the coupling with
perturbations at other scales cannot be ignored.  
The equation for evolution of density perturbations cannot be solved for
generic perturbations in the non-linear regime.  
N-Body simulations \cite{1998ARA&A..36..599B} 
are often used to study the evolution in this
regime, unless one requires only a limited amount of information and
quasi-linear approximation schemes \cite{1970A&A.....5...84Z,1989MNRAS.236..385G,1992MNRAS.259..437M,1993ApJ...418..570B,1994MNRAS.266..227B,1995PhR...262....1S,1996ApJ...471....1H,2002PhR...367....1B} or scaling 
relations
\cite{1977ApJS...34..425D,1991ApJ...374L...1H,1995MNRAS.276L..25J,2000ApJ...531...17Ka,1998ApJ...508L...5M,1994MNRAS.271..976N,1996ApJ...466..604P,1994MNRAS.267.1020P,1996MNRAS.278L..29P,1996MNRAS.280L..19P,2003MNRAS.341.1311S}
suffice.  

In N-Body simulations, we simulate a representative region of the
universe.
This representative region is a large but finite volume and periodic
boundary conditions are often used -- this is necessary considering
the the universe does not have a boundary. 
Typically the simulation volume is taken to be a cube.  
Effect of perturbations at scales smaller than the mass resolution of
the simulation, and of perturbations at scales larger than the box is
ignored.
Indeed, even perturbations at scales comparable to the box are under
sampled. 
It has been shown that for gravitational dynamics in an expanding
universe, perturbations at small scales do not influence collapse of
large scale perturbations in a significant manner
\cite{1974A&A....32..391P,1985ApJ...297..350P,1991MNRAS.253..295L,1997MNRAS.286.1023B,1998ApJ...497..499C}
as far as the correlation function or power spectrum at 
large scales are concerned.
Therefore we may assume that ignoring perturbations at scales much
smaller than the scales of interest does not affect results of N-Body
simulations. 
However, there may be other effects that are not completely understood
at the quantitative level \cite{2004astro.ph..8429B}. 

Perturbations at scales much larger than the simulation volume can affect the
results of N-Body simulations.  
Use of periodic boundary conditions implies that the average density in the
simulation box is same as the average density in the universe, in other words
we are assuming that there are no perturbations at the scale of the simulation
volume (or at larger scales). 
Therefore the size of the simulation volume should be chosen so that the
amplitude of fluctuations at that scale (and at larger scales) is ignorable. 
If the amplitude of perturbations at larger scales is not ignorable and the
simulations do not take the contribution of these scales into account then
clearly the simulation will not be a faithful representation of the model
being simulated. 
It is not obvious as to when fluctuations at larger scales can be considered
ignorable.
In this paper we will propose one method of quantifying this concept.
 
If the amplitude of density perturbations at the box scale is small
but not ignorable, simulations underestimate the correlation function though
the number density of small mass haloes does not change by much
\cite{1994ApJ...436..467G,1994ApJ...436..491G}. 
In other words, the formation of small haloes is not disturbed but their
distribution is affected by non-inclusion of long wave modes. 
The mass function of massive haloes changes significantly
\cite{1994ApJ...436..467G,1994ApJ...436..491G}. 
The void spectrum is also affected by finite size of the simulation volume if
perturbations at large scales are not ignorable \cite{1992ApJ...393..415K}. 

Significance of perturbations at large scales has been discussed in detail and 
a method (MAP) has been devised for incorporating the effects of these
perturbations \cite{1996ApJ...472...14T}.
These methods make use of the fact that if the box size is chosen to be large
enough then the contribution of larger scales can be incorporated by adding
displacements due to the larger scales independently of the evolution of the
system in an N-Body simulation. 
But this again brings up the issue of what is a large enough scale in any
given model such that these methods can be used to add the effect of larger
scales without introducing errors.  
Large scales contribute to displacements and velocities, and variations in
density due to these scales modify the rate of growth for small scales
perturbations \cite{1997MNRAS.286...38C}.  
The MAP algorithm \cite{1996ApJ...472...14T} can be improved by adding
corrections for this effect \cite{1997MNRAS.286...38C}.
The motivation for developing these tools was to enlarge the dynamic range
over which results of N-Body simulations are valid by adding corrections that
change the distribution of matter and velocities at scales comparable to the
simulation volume.

\begin{table}
\begin{center}
\begin{tabular}{||l|l|l|l|l||}
\hline
Name & $N_p$ & $L_{box}$ & $\epsilon$ & Cutoff \\
\hline
T\_300\_C\_0 & $256^3$ & $300$h$^{-1}$Mpc & $0.47$h$^{-1}$Mpc & No \\
T\_300\_C\_2 & $256^3$ & $300$h$^{-1}$Mpc & $0.47$h$^{-1}$Mpc & $k \leq 2 k_f$ \\
T\_300\_C\_3 & $256^3$ & $300$h$^{-1}$Mpc & $0.47$h$^{-1}$Mpc & $k \leq 3 k_f$ \\
T\_300\_C\_4 & $256^3$ & $300$h$^{-1}$Mpc & $0.47$h$^{-1}$Mpc & $k \leq 4 k_f$ \\
\hline
\end{tabular}
\caption{This table lists parameters of N-Body simulations we have used.  All
  the simulations were done using the TreePM code, configuration described in
  Bagla and Ray (2003).  We simulated the $\Lambda$CDM model with
  $\Omega_B = 0.05$, $\Omega_{dm}=0.3$, $k=0$, $h=0.7$ and $n=1$.  
  The first column lists name of the simulation, second column lists the
  number of particles in the simulation, the third column is the size of the
  box in physical units, fourth column lists the softening length for force
  and the last column lists the cutoff used in units of 
  the fundamental mode of the box $k_f$.  Phases were kept the same for the
  simulations listed here in order to make the comparison meaningful.} 
\end{center}
\end{table}

Our motivation in this work is to understand the effect of large scales on
scales that are much smaller than the simulation volume. 
We run a series of numerical experiments, these are described in \S{2}.
We discuss the proposed method of quantifying the effect of large scales in
\S{3}.  
Here, we propose a measure based on mass functions.
The minimum size of simulation volume for $\Lambda$CDM model as a function of
redshift is presented here.


\section{N-Body Simulations}

We carried out two series of N-Body experiments in order to study the effect
of perturbations at large scales on perturbations at small scales.  
The simulations were carried out using the TreePM method
\cite{2002JApA...23..185B,2003NewA....8..665B} and its parallel version
\cite{2004astro.ph..5220Ra}.  
We simulated the $\Lambda$CDM model with   $\Omega_B = 0.05$,
$\Omega_{dm}=0.3$, $\Omega_{tot}=1$, $h=0.7$ and $n=1$.  
In each of these series of simulations we ran a simulation without any
truncation of the power spectrum, save those imposed by the finite
size and resolution of the simulation box.
These simulations served as a reference for other simulations where we
truncated the power spectrum at small wave numbers. 
We used a sharp cutoff such that the power spectrum at $k \leq k_c$
was taken to be zero.  
A comparison of these simulations allows us to estimate the effect of
density perturbations at large scales on growth of perturbations at
small scales.
Detailed specifications of these simulations are listed in table~1
where the cutoff wave number is listed in the units of the fundamental
mode $k_f =  2 \pi / L_{box}$ of the simulation box. 

\begin{figure}
\includegraphics[width=3.3in]{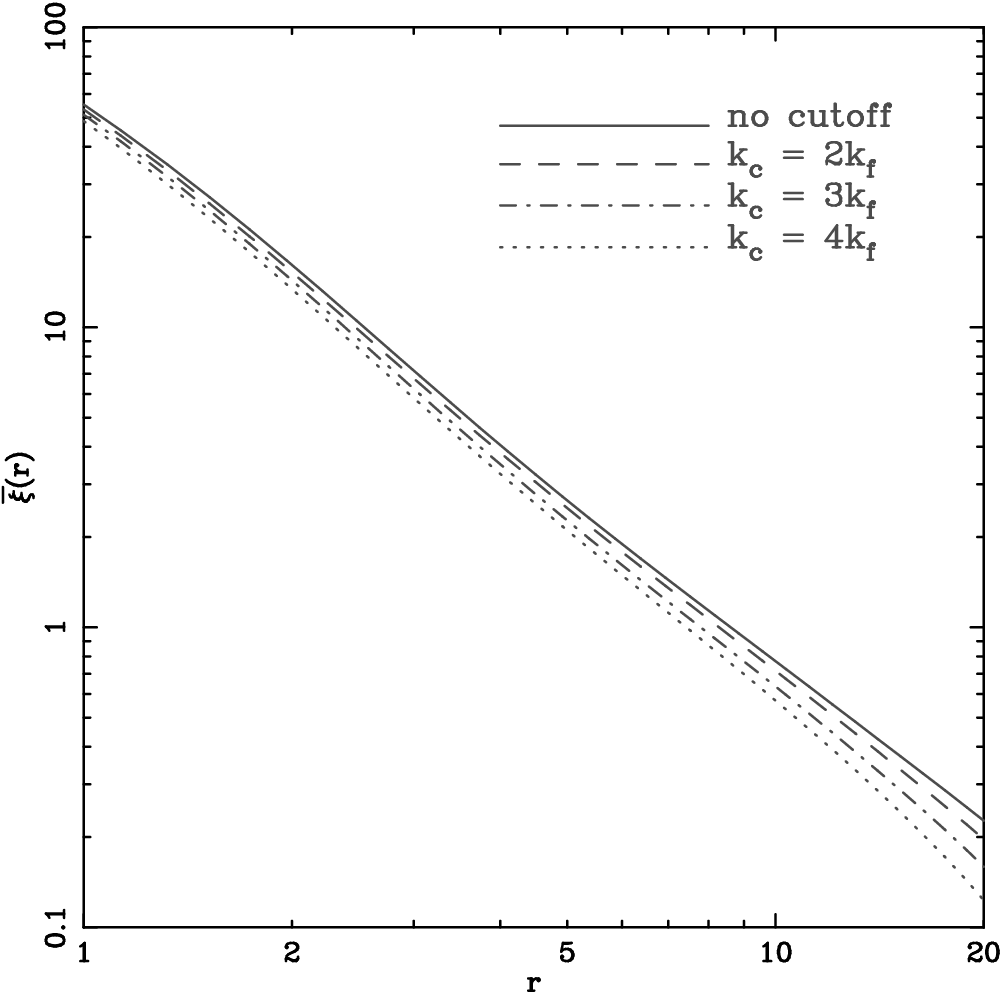}
\vspace{30pt}
\includegraphics[width=3.3in]{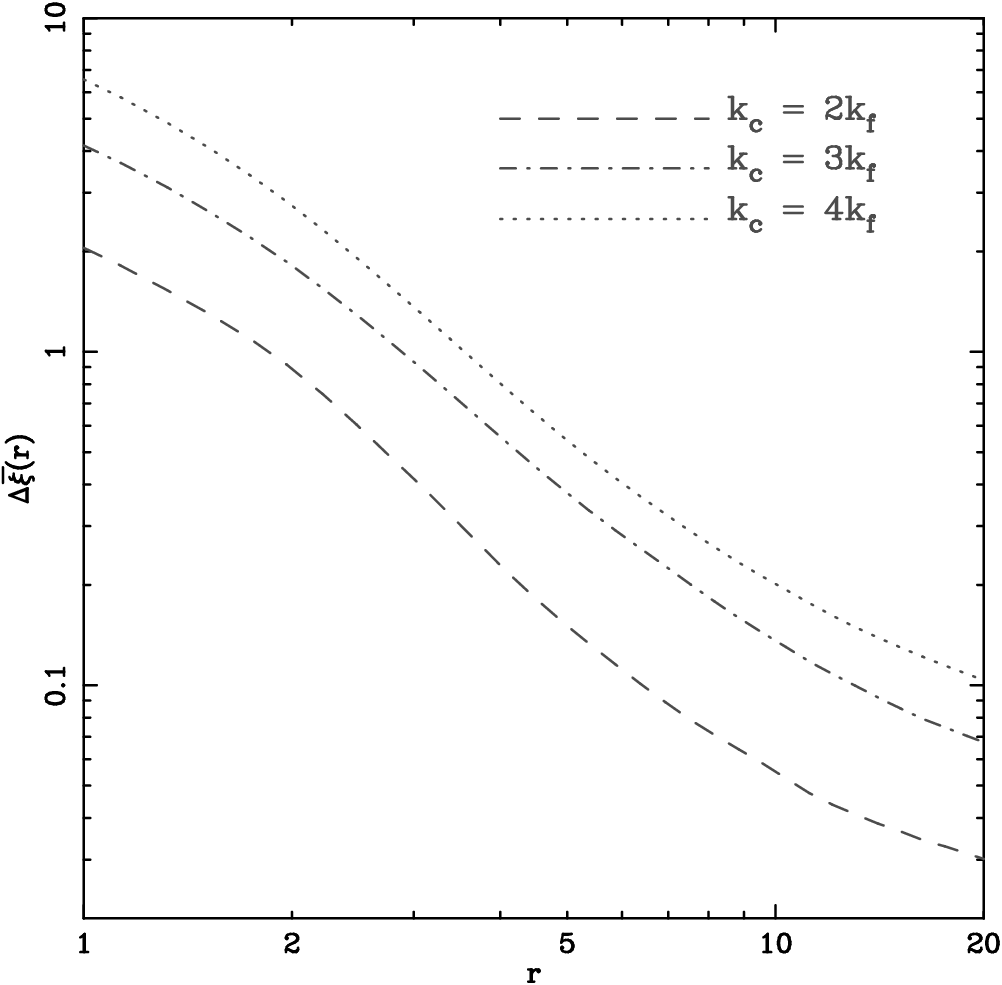}
\caption{The top panel shows the averaged correlation function $\bar\xi$ as a
  function of $r$ for the models listed in table~1.  The amplitude of the
  correlation function decreases as cutoff becomes smaller.  The lower panel
  shows the deviation of $\bar\xi$ for models with an explicit cutoff from the
  simulation T\_300\_C\_0.  The correlation function is underestimated by
  about $10\%$ at $r=2$h$^{-1}$Mpc for a cutoff of $100$h$^{-1}$Mpc.} 
\label{figcorr_cutoff}
\end{figure}

\begin{figure}
\includegraphics[width=3.3in]{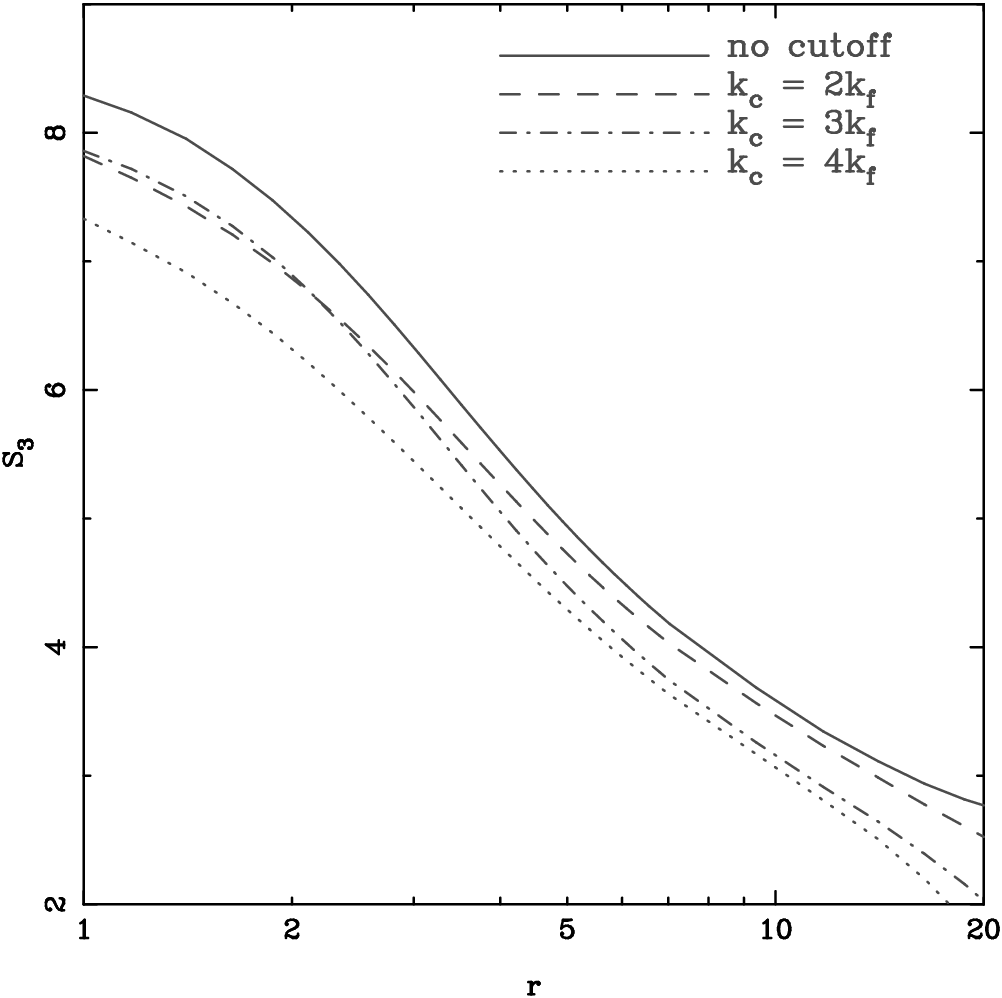}
\caption{This figure shows the skewness $S_3$ as a function of $r$ for the
  models listed in table~1.   The skewness decreases sharply as cutoff becomes
  smaller.}   
\label{figs3_cutoff}
\end{figure}

We compare the output of these simulations to see whether retaining or
dropping long wave modes affects quantities of interest at smaller scales.  
Nearly all the comparisons we carry out will concentrate on scales smaller
than $20$h$^{-1}$Mpc whereas the smallest cutoff we use is $75$h$^{-1}$Mpc,
so the two sets of scales are well separated in most cases we consider here.  
The amplitude of root mean square fluctuations in mass at $r=75$h$^{-1}$Mpc
and $z=0$ is $\sigma \sim 0.05$ in the model considered here and we can
certainly consider this to be small. 

Figure~\ref{figcorr_cutoff} shows the averaged correlation function $\bar\xi$
as a function of $r$ at $z=0$ for the models listed in table~1.   
The amplitude of the correlation function decreases as cutoff becomes
smaller. 
The shape of the correlation function does not change at small scales.
This result follows expectations and is indeed similar to figure~2 in
\cite{1994ApJ...436..491G}. 
Thus the overall effect of perturbations at very large scales is to enhance
the amplitude of fluctuations at smaller scales.
Ignoring larger scales leads to an underestimate of correlation function at
small scales. 
The correlation function is underestimated by $10\%$ at $r=2$h$^{-1}$Mpc for a
cutoff of $100$h$^{-1}$Mpc.
At this scale, $\sigma=0.04$ for $z=0$ in the model we have simulated.
If we were to compute the linearly extrapolated $\bar\xi$ then the difference
between the model with and without cutoff is much smaller than $10\%$. 
We have plotted the skewness $S_3$ as a function of scale in
figure~\ref{figs3_cutoff} for the same set of models. 
The difference in models here is significant.  
It is clear from these two figures that fairly large scales ($l \leq
100$h$^{-1}$Mpc) make a significant contribution to clustering at small scales
($l \leq 10$h$^{-1}$Mpc).

It is possible to correct for the contribution of larger scales for moments
of particle distribution \cite{1994A&A...281..301C}.
Thus if we are interested only in the moments then we can circumvent this
problem and compute the correct answer.
There are many quantities of interest other than the moments of distribution
and there is no generic method for computing the corrections and take the
effect of large scales into account.

A method has been devised to incorporate the effect of displacements and
velocities contributed by large scales \cite{1996ApJ...472...14T}.
Here, displacements and velocities contributed by larger scales are computed
using the Zel'dovich approximation \cite{1970A&A.....5...84Z} and added to the
numbers computed in the N-Body simulations. 
Essentially this assumes that there is no coupling in velocity contributed by
modes within the box and modes larger than the box.  
With this assumption, other methods for computing the displacement due to
modes outside the box can also be used in place of the Zel'dovich
approximation. 
The assumption of no mode coupling is true only in the linear regime and hence
we still require the amplitude of perturbations at the scale of the simulation
box to be much smaller than unity. 
How small the amplitude should be can only be determined by trial and error.
It has been pointed out that the effect of mode coupling can be emulated by
enhancing the displacements without modifying the velocities
\cite{1997MNRAS.286...38C}. 
The amount of enhancement needed cannot be derived from first principles. 
These tools essentially enhance the dynamic range on an N-Body simulation by
making these corrections, displacements due to large scales correct the
velocity fields at scales comparable to the simulation volume. 

\begin{figure*}
\includegraphics[width=6.5in]{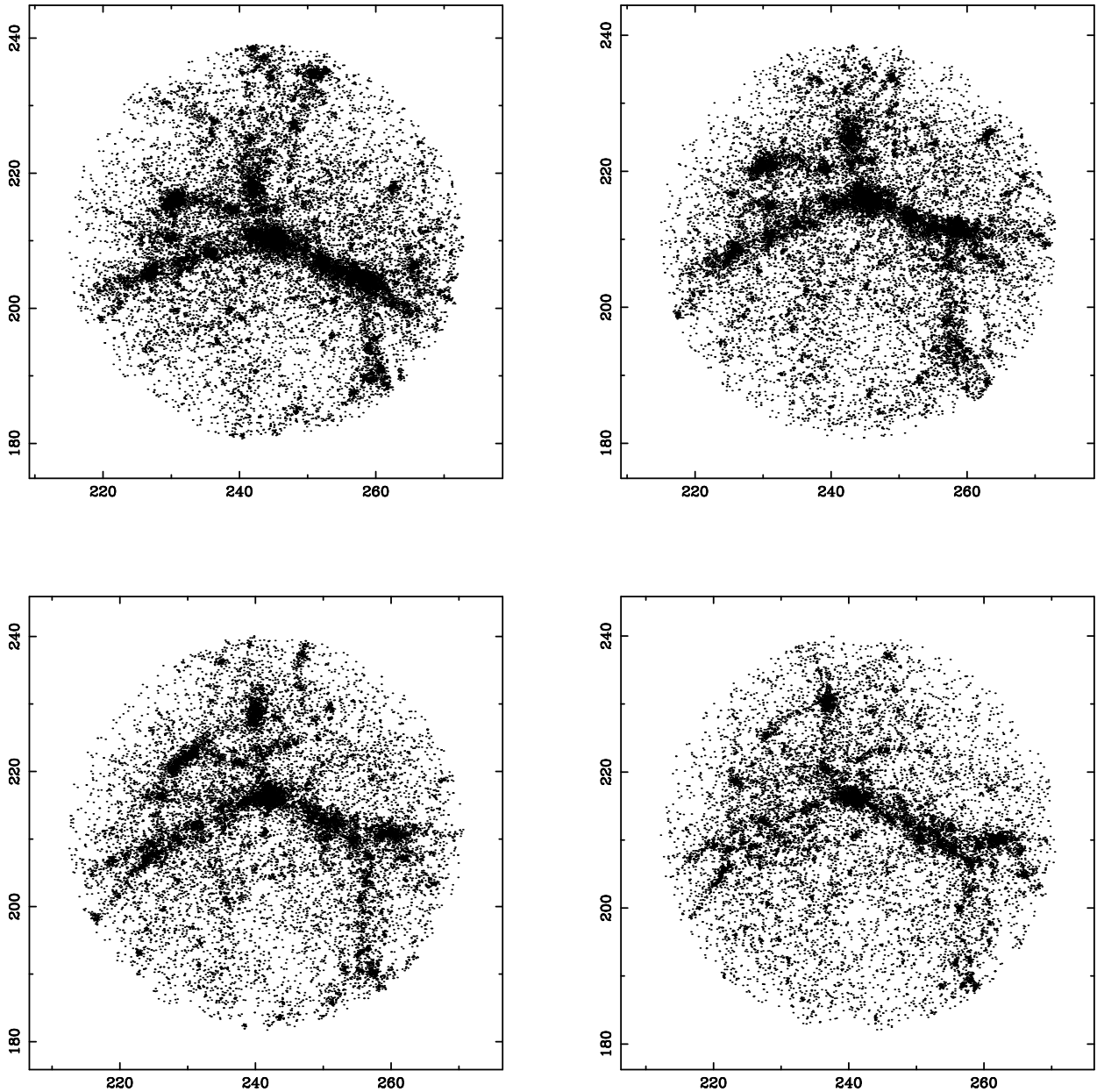}
\caption{Four panels of this figure show the projected particle distribution
  in the neighbourhood of a massive cluster from simulations of the
  $\Lambda$CDM model.  The top left panel is for the first simulation in
  table~1, top right panel is for the second simulation, lower left panel is
  for the third simulation and the lower right panel is for the fourth
  simulation in the table.  The total mass in the central massive halo
  decreases as the large scale cutoff is introduced and then decreased.  The
  number, masses and the distribution of smaller clumps around the central
  clump also change significantly as the cutoff is reduced below
  $150$h$^{-1}$Mpc.}
\label{fig_part}
\end{figure*}

Corrections to displacements and velocities can be made without worrying about
mode coupling if their effect is small in some absolute sense. 
An important fact to keep in mind is that none of these methods can be
effective if the displacements contributed by large scale modes move
two collapsed objects to the same location, or move matter that has fallen
into a collapsed structure out of it.  
If the contribution of modes that can affect collapsed structures is
not taken into account then the properties of these objects, e.g. mass,
angular momentum, density profile, etc., may differ significantly from their
asymptotic values. 
This is illustrated in figure~\ref{fig_part}. 
This figure shows the projected particle distribution in the neighbourhood of
a massive cluster from simulations of the $\Lambda$CDM model. 
Four panels correspond to the models listed in table~1.
The total mass in the central massive halo decreases as the large scale cutoff
is introduced and then decreased.  
The number, masses and the distribution of smaller clumps around the central
clump also change significantly as the cutoff is reduced below
$150$h$^{-1}$Mpc. 

Figure~\ref{fig_massfn} shows the fraction of mass in collapsed structures
with mass greater than $M$ for the models listed in table~1.
This fraction is plotted as a function of mass $M$. 
The solid curve shows $F(>M)$ for the simulation without an explicit cutoff,
of course there is an implicit cutoff at the box size ($300$h$^{-1}$Mpc).
Other curves mark the function for different values of the cutoff and as the
cutoff scale becomes smaller $F(>M)$ decreases at large $M$. 
At smaller masses ($M \leq 10^{14}$M$_\odot$), the difference in models is
negligible. 
Difference between different curves at the high mass ($M \simeq
10^{15}$M$_\odot$) end is significant, more than a factor two between the
extreme curves.  
Thus ignoring large scale modes results in an underestimate of the number of
massive haloes. 
$F(>M)$ varies rapidly up to a cutoff of $150$h$^{-1}$Mpc and changes very
little as the cutoff moves to larger scales.  
We may conclude from here that scales larger than $150$h$^{-1}$Mpc do not
contribute significantly to collapsed structures in the currently favoured
$\Lambda$CDM models. 

\section{The Proposed Criterion}

In the previous section we have argued that the non-trivial contribution of
large scale modes is the one that leads to collapse of haloes, other effects
like displacements can be incorporated, in principle, using algorithms like
MAP \cite{1996ApJ...472...14T,1997MNRAS.286...38C}. 
We can use this to devise a criterion to decide whether the box size of a
simulation is sufficiently large or not.

The collapsed mass fraction in haloes of mass $M$ or larger is given by
\cite{1974ApJ...187..425P,1991ApJ...379..440B}
\begin{equation}
F(M,z) = {\rm erfc}\left[ \frac{\delta_c}{\sqrt{2} \sigma_L\left(M,z\right)}
\right] 
\label{coll_mass}
\end{equation}
The parameter $\delta_c$ indicates the linearly extrapolated density contrast
at which the perturbation is expected to collapse and virialise in non-linear
spherical collapse, its value is taken to be $1.686$ here.  
The precise value of this parameter is not very relevant here.
Here $\sigma(M,z)$ is the root mean square (rms) amplitude of fluctuations at
the mass scale $M$ and redshift $z$.  
\begin{eqnarray}
\sigma_L^2(M,z) &=& 9 D_{+}^2(z) \int\limits_0^\infty \frac{dk}{k} \frac{k^3
  P(k)}{2 \pi^2} \left[\frac{\sin{k r} - k r \cos{k r}} {k^3 r^3} \right]^2
   \nonumber \\
M &=& \frac{4 \pi}{3} {\bar\rho} r^3 
\label{sig_int}
\end{eqnarray}
Here $P(k)$ is the power spectrum of density fluctuations, linearly
extrapolated to $z=0$ and $D_+(z)$ is the growing mode in the linear
perturbation theory normalised so that $D_+(z=0)=1$.  
The average matter density in the universe is denoted by $\bar\rho$. 

In an N-Body simulation, the initial conditions sample a range of values of
wave number $k$. 
The amplitude of {\it rms} fluctuations in the simulation will be different
due to partial sampling of modes. 
The periodic boundary conditions restrict us to a discrete set of wave numbers
in the range of scales studied and sampling is partial due to sparse sampling
of $k-$space at scales comparable to the size of the simulation volume.
Wave numbers smaller than the fundamental mode of the simulation volume are
not sampled, nor are wave numbers larger than the Nyquist frequency of the
mesh used for generating initial conditions. 
From the form of the integral it is clear that at a given scale $r$, wave
modes with $k \leq 2\pi /r$ contribute more significantly than the rest.  
Given this and the fact that most modern N-Body simulations have sufficient
dynamic range, we can concentrate on the lower limit of the range of wave
numbers sampled in an N-Body simulation as modes with $k \geq k_{max}$ do not 
influence scales resolved in the simulation in any significant manner. 
We can then estimate the {\it rms} fluctuations in N-Body simulations by
changing the lower limit of the integral in eqn.(\ref{sig_int}) from $0$ to $2
\pi /L_{box}$ while leaving the upper limit unchanged. 
The fluctuations will now be a function of the cutoff as well:
\begin{eqnarray}
& & \sigma_L^2(M,z,L_{box}) = 9 D_{+}^2(z) \nonumber \\
& & \;\;\;\;\;\;\;\;\;\; \times \int\limits_{2\pi/L_{box}}^\infty
  \frac{dk}{k} \frac{k^3 P(k)}{2 \pi^2} \left[\frac{\sin{k r} - k r \cos{k r}}
  {k^3 r^3} \right]^2 
\end{eqnarray}
We can use this in eqn.(\ref{coll_mass}) and obtain $F(M,z,L_{box})$, the
expected collapsed mass fraction in an N-Body simulation. 
Here we have approximated the sum over the discrete set of $k-$modes by an
integral and assumed that the upper limit on the wave numbers sampled is less
relevant for estimating the mass function at scales of interest. 

\begin{figure}
\includegraphics[width=3.3in]{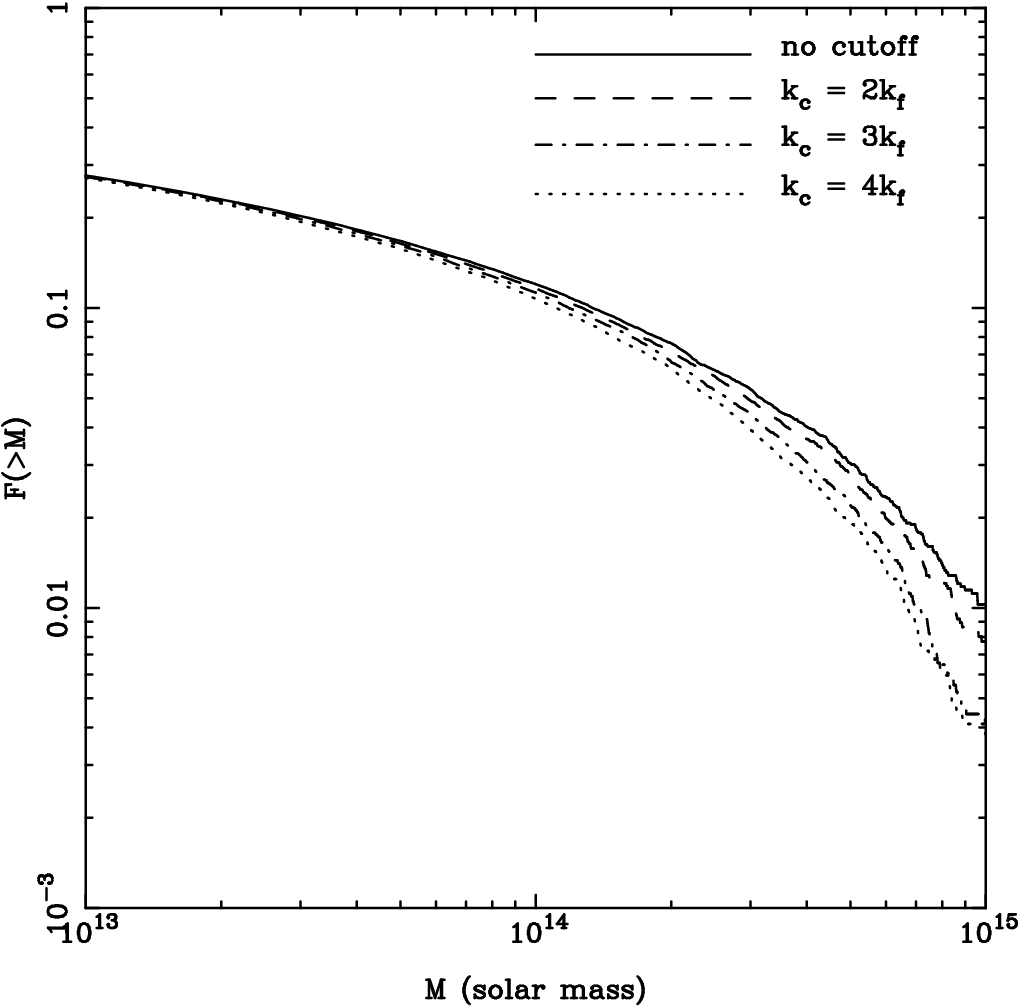}
\caption{This figure shows the fraction of mass in collapsed structures with
  mass greater than $M$ for the models listed in table~1.  While the change in
  cutoff does not change the collapsed mass fraction for $M \leq
  10^{14}$M$_\odot$, the collapsed fraction is underestimated at larger masses
  when we remove the large scale modes.  Suppression of mass function around
  $M=10^{15}$M$_\odot$ is significant when the cutoff is at scales smaller
  than $150$h$^{-1}$Mpc.}  
\label{fig_massfn}
\end{figure}

\begin{figure}
\includegraphics[width=3.3in]{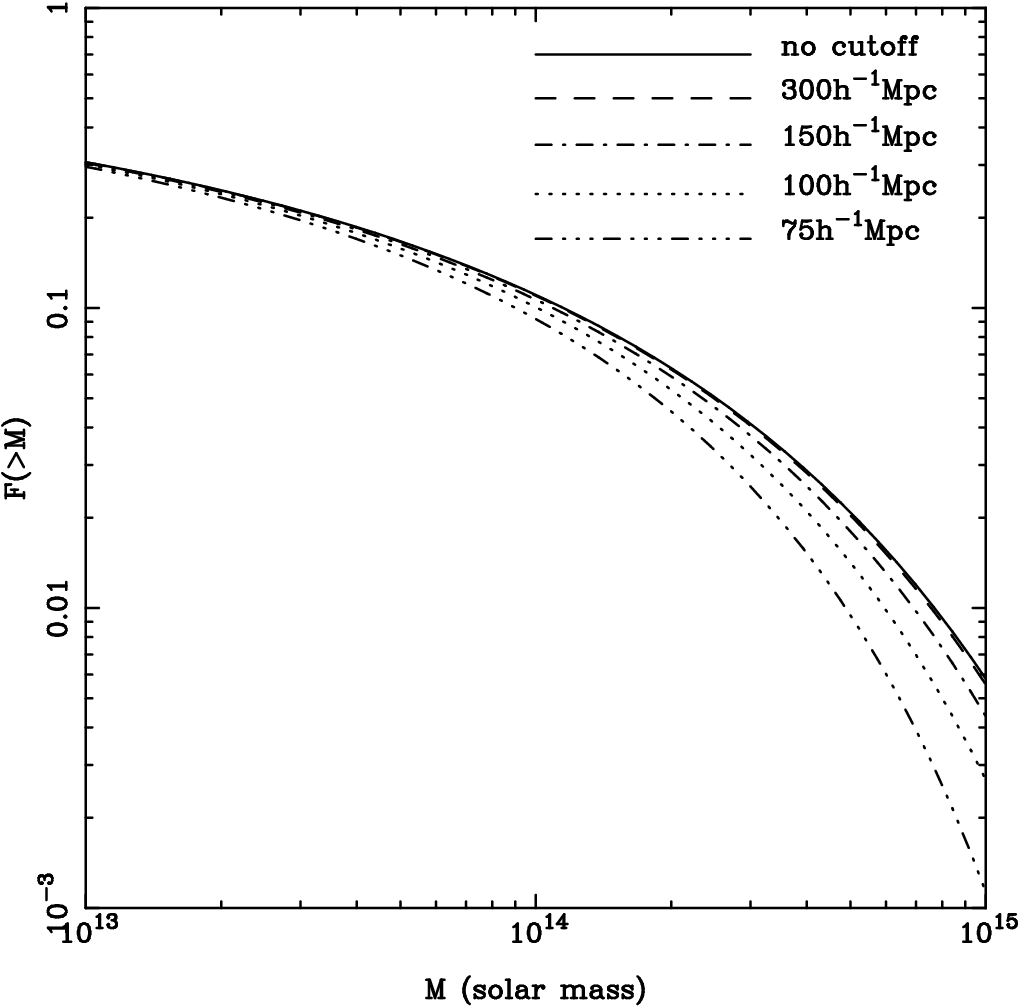}
\caption{This figure shows the fraction of mass in collapsed structures with
  mass greater than $M$ for the $\Lambda$CDM model used here.  This was
  computed using the Press-Schechter mass function (Press \& Schechter, 1974).
  A cutoff was used to remove contributions from large scales (see
  \S{3} for details).  Note that these values of cutoff
  coincide with those used in simulations of models listed in table~1 and
  plotted in figure~4.} 
\label{fig_psmassfn}
\end{figure}

In the previous section we found that in N-Body simulations the collapsed mass
fraction does not change much if the cutoff is larger than $150$h$^{-1}$Mpc. 
This conclusion is reaffirmed by the theoretical calculation of the mass
function and collapsed mass fraction. 
Figure~\ref{fig_psmassfn} shows $F(M,L_{box})$ as a function of mass $M$ for
different values of the cutoff scale $L_{box}$.  
The mass in the most massive collapsed structures changes rapidly with the
cutoff, implying that the number density of most massive structures depends
strongly on the large scale modes.
Comparison with figure~\ref{fig_massfn} shows that the theoretical calculation
and simulations give comparable results. 

The criterion we propose here is this: the physical scale $L_{box}$
corresponding to the size of the simulation volume can be considered to be
large enough if the expected fraction of mass in haloes $F(M,z,L_{box})$ is
comparable to $F(M,z)$, the fraction of mass in haloes when the full
spectrum is taken into account. 
In other words, we require convergence of expected mass in haloes for the 
simulation volume to be considered large enough so that all the relevant
scales are contained within it.  
As before, we are interested in the effect of large scales on scales much
smaller than $L_{box}$. 
Therefore we wish to see this convergence at mass scales of typical haloes. 
We define the mass scale of non-linearity as $\sigma_L(M_*,z) = 1$ and study
the convergence of mass function at this scale and at neighbouring mass
scales. 

\begin{figure}
\includegraphics[width=3.3in]{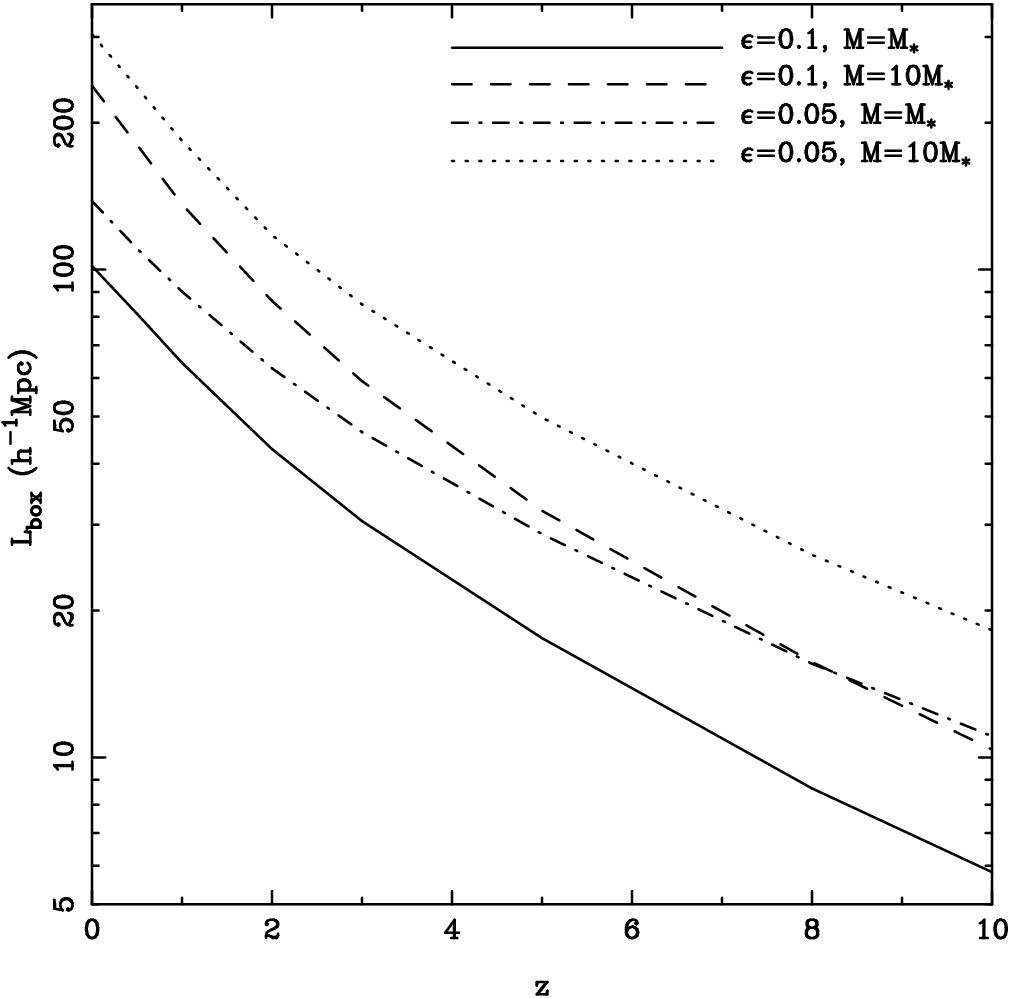}
\caption{We have plotted $L_{box}$ as a function of redshift $z$ for the
  $\Lambda$CDM model (see \S{2} for the values of cosmological parameters).  
  Using a smaller box size than outlined here leads to a greater underestimate
  of the collapsed mass in haloes than given by conditions used for these
  curves.}    
\label{fig_lcdm}
\end{figure}

We study two masses for the $\Lambda$CDM model: $M_*$ and $10 M_*$.
As we shall see, these are more relevant than smaller mass scales in most
situations. 
We require $F(M,z,L_{box})= (1 - \epsilon) F(M,z)$ to find the
threshold length scale $L_{box}$, with $\epsilon=0.05$ and a less
conservative limit of $\epsilon=0.1$.
These criteria allow us to develop a feel for this approach.  
We certainly require a reasonable convergence at the scale of
non-linearity in simulations. 
For some applications, as when studying rich clusters of galaxies, we require
good convergence for very massive haloes. 
Figure~\ref{fig_lcdm} shows $L_{box}$ as a function of redshift $z$
according to these criteria for the $\Lambda$CDM model (see \S{2} for
the values of cosmological parameters). 
The solid curve ($\epsilon=0.1$) and the dot-dashed curve
($\epsilon=0.05$) are for $M=M_*$, the dashed curve ($\epsilon=0.1$)
and the dotted curve ($\epsilon=0.05$) are for $M=10 M_*$.
For the $\Lambda$CDM model, $M_*(z=0)=1.2 \times 10^{14}$M$_\odot$
therefore the criteria used here refer to mass scales of typical
clusters of galaxies and rich clusters of galaxies, respectively,
at redshift $z=0$.  
At $z=0$, it is clear that a box larger than $100$h$^{-1}$Mpc is
required even if we are interested in haloes with mass $M_*$ and can
tolerate an offset of $10\%$ in collapsed mass. 
Using a smaller box size leads to a greater underestimate of the collapsed
mass in haloes.   
The requirement becomes more stringent if we wish to study rich
clusters of galaxies or use a tolerance level of $5\%$
($\epsilon=0.05$). 

At $z=3$,  $M_* =1.2 \times 10^{11}$M$_\odot$.  
This is the mass of a typical galaxy halo at that redshift and to
get the statistics of these objects right we should have a box size of
at least $30$h$^{-1}$Mpc ($\epsilon=0.1$), the required box size
increases to $50$h$^{-1}$Mpc if we require $\epsilon=0.05$ instead. 
This, or an even larger simulation box should be used if we wish to
study the inter-galactic medium as there will be more matter left in
uncollapsed regions if we use a smaller simulation box. 
Bright Lyman break galaxies are likely to be in more massive than typical
haloes and a box size of $60$h$^{-1}$Mpc or larger is needed to study
these in an N-Body simulation. 

At higher redshifts, simulations are often used for studying
reionisation of the universe. 
At $z=10$,  $M_* =2 \times 10^6$M$_\odot$.
Sources of ionising radiation are likely to reside in much more
massive haloes and hence we should use a simulation box that is at
least $20$h$^{-1}$Mpc across ($M=10 M_*$ and $\epsilon=0.05$).
If we relax the requirement to $\epsilon=0.1$ then the simulation
volume should be more than $10$h$^{-1}$Mpc across.
Of course, very high peaks that may be the first sources of ionising radiation
are rare and a very large simulation box is required to study patchiness in
ionising radiation and hence reionisation \cite{2004ApJ...609..474B}.

Clearly, the requirement that the mass in collapsed haloes should not depend
significantly on scales larger than the simulation box is fairly stringent and
in some cases it may make it difficult to address the physical problem of
interest. 
This restriction is less stringent than requiring that the correlation
function of haloes converge as it has been shown that the mass function
converges well before the correlation function of haloes
\cite{1994ApJ...436..467G,1994ApJ...436..491G}. 
However, if the mass function has converged then tools like MAP
\cite{1996ApJ...472...14T} can be used to obtain the correct distribution of
haloes. 
Of course, curves can be drawn with different requirements on the convergence
of collapsed mass as long as these requirements are in consonance with the
physical application for which the simulation is being run.

\begin{figure}
\includegraphics[width=3.3in]{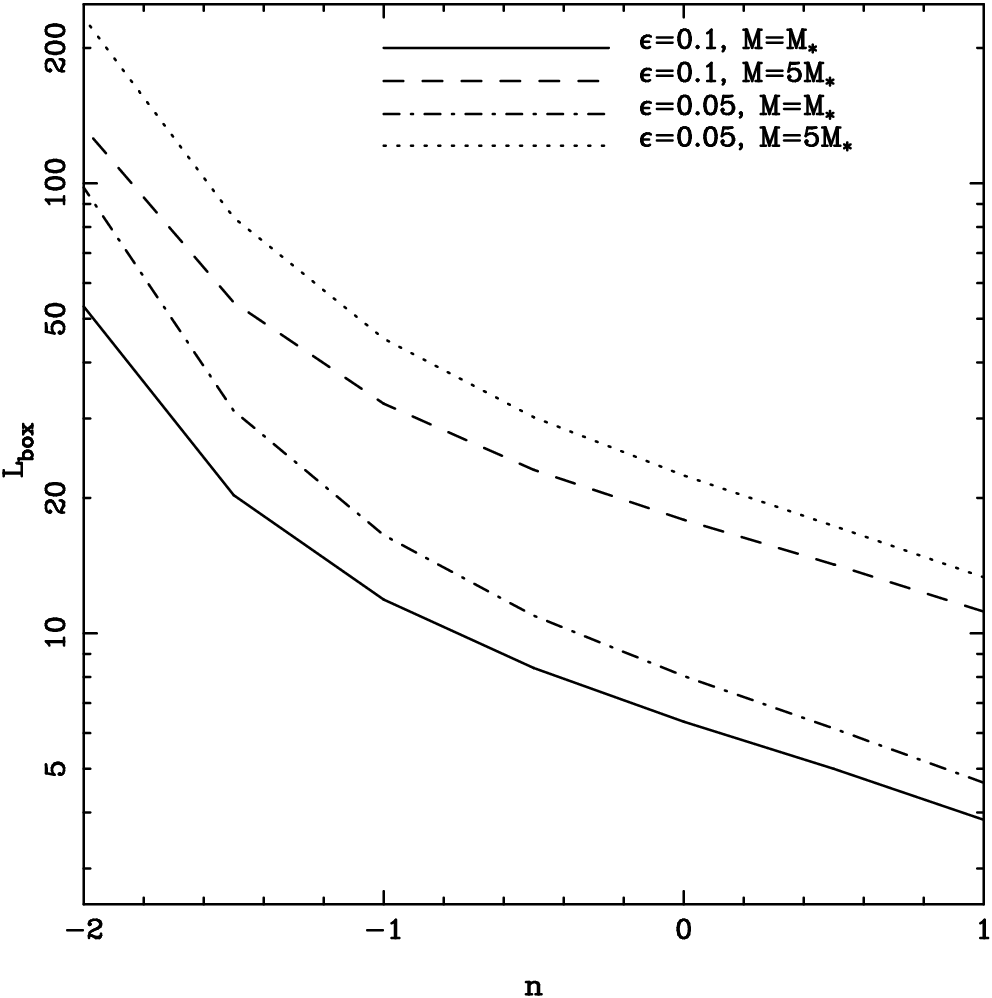}
\caption{$L_{box}$ is shown as a function of the index of power spectrum for
  power law models, $P(k) = A k^n$.  The power spectrum is normalised such
  that $\sigma_L(r=1) = 1$.  We have assumed an Einstein-de Sitter
  background for these models.  It is clear from this figure that it is
  difficult to explore the highly non-linear regime for smaller values of
  $n$.}  
\label{fig_index}
\end{figure}

The curves for $\epsilon=0.1$ are not parallel to the curves for
$\epsilon=0.05$, this is due to a dependence on the shape of the power
spectrum. 
This dependence is brought out very clearly in figure~\ref{fig_index}. 
Here we show $L_{box}$ as a function of the index of power spectrum for power
law models, $P(k) = A k^n$.
The power spectrum is normalised such that $\sigma_L(r=1) = 1$. 
We have assumed an Einstein-de Sitter background for these models. 
The solid curve ($\epsilon=0.1$) and the dot-dashed curve
($\epsilon=0.05$) are for $M=M_*$, the dashed curve ($\epsilon=0.1$)
and the dotted curve ($\epsilon=0.05$) are for $M=5 M_*$.
It is clear from this figure that it is difficult to explore the highly
non-linear regime for smaller values of $n$, required box size increases
rapidly as $(n+3) \rightarrow 0$.

These figures show that the convergence of collapsed mass in
haloes happens slowly for most models of interest.
The offset in collapsed mass for the the simulated model (with a finite
sampling of scales) from the collapsed mass for the model that we wish to 
simulate is a useful indicator of the relevance of large scales. 
This feature allows us to use the criterion to decide the size of the
simulation volume; the criterion supplies the lower bound on the size.  


\section{Conclusions}

We have studied the influence of long wave modes on gravitational
clustering at small scales.  
We find that for the $\Lambda$CDM model, scales larger than $100$h$^{-1}$Mpc
affect the mass function of haloes and the distribution of matter at scales as
small as a few Mpc. 
The effect of long wave modes not present in an N-Body simulation can be
incorporated independently of the evolution at small scales for some
quantities but making such corrections is not possible in general.
In particular, it is not possible to make corrections if the contribution of
large scales changes the mass of collapsed haloes by a significant amount. 
We can turn this argument around and check whether a given size of the
simulation volume will give (close to) correct results for total
collapsed mass in haloes or not.
This can be done using simple analytical formulae outlined in \S{3}. 
The fractional deviation of collapsed mass from its expected value if
density fluctuations at all scales are taken into account is a good indicator
of the influence of large scales. 
The fractional deviation ($\epsilon$) can be checked at a given mass scale
$M$ of physical interest.   
Given a choice of $M$, $\epsilon$ and the redshift $z$ at which output
of the simulation is to be studied, we can compute the recommended
minimum size for the simulation volume.  
It is important to note that this scale is the minimum required and a larger
simulation may be warranted by other requirements.

The mass in collapsed haloes converges faster than the correlation function of
haloes, implying that an even larger simulation volume may be required.  
Corrections to positions and velocities due to large scales can be made using
tools such as MAP \cite{1996ApJ...472...14T,1997MNRAS.286...38C}. 
Thus we can choose the box size by requiring convergence of mass in haloes and
then use MAP to get the correct distribution and velocity field at all
scales. 
Of course, if MAP is not being used then a larger box size than the minimum
indicated by convergence of mass function may be required. 
Of course, using a large box size can make it difficult to address interesting
questions using N-Body simulations. 
We believe that it is better to be cautious rather than obtain unreliable
answers, unless one can make a convincing case that the relevant physical
quantities are not as sensitive as mass in collapsed haloes to the size of the
simulation box.


\section*{Acknowledgements}

Numerical experiments for this study were carried out at cluster computing
facility in the Harish-Chandra Research Institute
(http://cluster.mri.ernet.in).  
This research has made use of NASA's Astrophysics Data System. 
We thank the anonymous referee for comments and useful suggestions.



\label{lastpage}

\end{document}